\pdfoutput=1

\documentclass[12pt,letterpaper]{article}
\usepackage{jheppub}
\usepackage{journals}
\usepackage{graphicx}
\usepackage{amssymb}
\usepackage{placeins}
\usepackage{jheppub}
\usepackage{journals}
\usepackage{graphicx}
\usepackage{amsmath}
\usepackage{placeins}
\usepackage{wrapfig}
\usepackage{subfig}
\usepackage{amssymb}
\usepackage{hyperref}
\usepackage[toc,page]{appendix}
\usepackage{tikz}

\newcommand\bl{\color{blue}}

\begin		{document}

\title		{Holographic Kolmogorov-Sinai entropy and the quantum Lyapunov spectrum}

\author[a]{Georg Maier,}
\author[a]{Andreas Sch\"afer,}
\author[b,c]{Sebastian Waeber}

\affiliation[a] {Institute of Theoretical Physics, University of Regensburg, D-93040 Regensburg, Germany}
\affiliation[b] {Department of Physics, Technion, Haifa 32000, Israel}
\affiliation[c] {Department of Physics, University of Washington, Seattle WA 98195-1560, USA
}
		\emailAdd	{georg.maier@physik.uni-regensburg.de}
		\emailAdd	{andreas.schaefer@physik.uni-regensburg.de}
		\emailAdd	{swaebe@uw.edu}

\abstract
{In classical chaotic systems the  entropy, averaged over initial phase space distributions, follows an universal behavior. While approaching thermal equilibrium it passes through a stage where  it grows linearly, while the growth rate, the Kolmogorov-Sinai entropy, is given by the sum over all positive Lyapunov exponents. A natural question is whether a similar relation is valid for quantum systems. We argue that the Maldacena-Shenker-Stanford bound on quantum Lyapunov exponents $\lambda$ implies that the upper bound on the growth rate of the entropy, averaged over states in Hilbert space that evolve towards a thermal state with temperature $T$ and entropy $S_{eq}$, should be given by $S_{eq} \pi T =\sum_{\lambda >0}2 \pi T$.  Strongly coupled, large $N$ theories with black hole duals should saturate the bound. By studying a large number of isotropization processes  of random, spatially homogeneous, far from equilibrium initial states in large $N$, $\mathcal{N}=4$ Super Yang Mills theory at strong coupling and computing the ensemble averaged growth rate of the dual black hole's apparent horizon area, we find both an analogous behavior  as in classical chaotic systems and numerical evidence that the conjectured bound on averaged entropy growth is saturated granted that the  Lyapunov exponents  are degenerate $\lambda = \pm 2 \pi T$. This  fits to the behavior of classical systems with plus/minus symmetric Lyapunov spectra, a symmetry which implies the validity of Liouville's theorem. \iffalse, suggesting that the latter might also apply to $N\rightarrow \infty$ strongly coupled gauge theories.\fi}
\maketitle
\section{Introduction}
\label{intro}
A quantum mechanical description of chaotic many body systems is of interest for a multitude of research areas in physics, especially in the context of condensed matter physics, thermalization and quantum information theory. In classical physics the question of "how chaotic" a system is, can be quantified by examining the rate with which phase space trajectories $X_i(t)$ diverge from one another,  if their starting points are separated by an infinitesimal perturbation of initial conditions $\delta X_j(0)$. In chaotic systems the separation of initially almost identical paths in phase space has an exponential behavior and the singular values  of the matrix $\delta X_i(t)/\delta X_j(0)$ grow or contract as  $e^{\lambda t}$. In the late time limit $t \to  \infty$ the exponents represented by $\lambda$ are referred to as Lyapunov exponents. In quantum theories this behavior is encoded in  out-of-time-order correlators (OTOCs) \cite{gor}. The  quantum Lyapunov exponents can  be extracted from the exponential growth rate of OTOCs $\sim e^{\lambda_{\text{OTOC}} t}$ at late times $\frac{1}{T}\ll t$. \\ \indent In recent years OTOCs and these exponents, encoding the speed with which quantum systems scramble information, received much attention \cite{jens, brown, key, brown2, blake, blake2, groz}, especially after it was shown that there exists an upper bound \cite{She} for $\lambda_{\text{OTOC}} \leq 2 \pi T$ and thus an upper bound on the speed of the development of quantum chaos. Furthermore, systems that are holographic duals to Einstein gravity have been found to saturate this bound \cite{She, She2, She3, She4}.  However, to the best of our knowledge, there exists no formal proof as to exactly which quantum systems show this behavior for which operators. Therefore, we focus an the best established holographic system, namely $AdS_5/CFT_4$, where this bound is known to be saturated. \\
\indent For classical  chaotic systems the Kolmogorov-Sinai entropy  which, contradicting its name, is not an entropy, but rather an entropy  growth rate, provides information about the Lyapunov spectrum at time scales that are relevant for thermalization or dissipation $t \sim \frac{1}{T}$. The heuristic idea is that, starting from some initial  ensemble of configurations in phase space, whose evolution is described by a chaotic, Hamiltonian system, will lead to a fractal shaped deformation of  the phase space volume of this ensemble, requiring more and more  phase space cells (used to evaluate the coarse grained entropy $S$) to cover its shape, while Liouville's theorem ensures that its  volume stays unchanged. Those directions in phase space for which the Lyapunov exponents are positive, will contribute to the growth of the number of  needed cells and thus contribute to (coarse grained) entropy growth\footnote{It is only the coarse grained entropy that grows, the fine grained entropy stays constant.}, such that naively we could have
\begin{equation}
\frac{d S}{dt} = \sum_{\lambda > 0}\lambda.
\label{entr_gen1}
\end{equation}
The growth rate $\frac{d S}{dt} = S_{KS}$ is the Kolmogorov-Sinai entropy. Empirically we know that this relation between entropy growth and Lyapunov exponents of chaotic systems is only correct with further specifications.  To begin with for a thermalizing system this statement can clearly only be true for some intermediate time period before the system reaches thermal equilibrium. Moreover the right hand side of the above equation is independent of initial conditions, while the time derivative of the growing entropy will depend on the initial state. Thus, in general only the ensemble averaged entropy, where we average over a large ensemble of initial phase space configurations that are far from equilibrium, will allow us to determine the sum over all positive Lyapunov exponents, as demonstrated in \cite{bar}.\\
\indent In \cite{Han} the authors speculated that  the Maldacena-Shenker-Stanford (MSS) bound implies an upper bound on entropy growth, which should be saturated for conformal theories with Einstein gravity duals in the bulk, i.e. that {\it black holes are also the fastest entropy generators with $dS/dt= \sum_{\lambda>0} 2 \pi T$}. In this contribution we propose a slightly modified version of this conjecture, guided by observations made in classical statistical mechanics \cite{bar}. We are going to argue  that strongly coupled, large $N$ theories, holographically dual to Einstein gravity, fulfill 
\begin{equation}
\Big  \langle \frac{d S}{dt} \Big \rangle = \sum_{\lambda > 0} 2 \pi T,
\label{entr_gen}
\end{equation}
where $\langle \cdot \rangle$ denotes the Hilbert space average over states that initially are far from equilibrium\footnote{To be more precise with this we mean states whose initial entropy is only a few percent of the equilibrium entropy.} and evolve towards the same thermal state with temperature $T$. Relation (\ref{entr_gen}) is supposed to hold until thermal equilibrium is almost reached, at which point $\langle \frac{d S}{dt}  \rangle$ gradually decreases to $0$, in analogy to classical chaotic systems. We examine this numerically by studying, via holography, far-from-equilibrium istoropization in $\mathcal{N}=4$ Super Yang Mills theory (SYM). We determine the number of Lyapunov exponents from the number of degrees of freedom of the equilibrated black hole, which is taken to be (albeit in all generality not proven to be) the same as its Bekenstein-Hawking entropy $S_{eq}$. \\ \indent For a (semi-)classical\footnote{In section \ref{class} we classify in what sense this calculation is semi-classical.} Yang-Mills theory the Lyapunov spectrum computed on the lattice can be split into three parts of which each belongs to one third of the degrees of freedom\footnote{Here we include also the unphysical ones that can be removed via gauge symmetry together with the physical ones into the common term 'degrees of freedom'. The discussion of our lattice calculation in section \ref{class} clarifies why this is done.}. One third of the Lyapunov exponents $\lambda^+_i$ is positive and their sum is the Kolmogorov-Sinai entropy, one third is negative with $\lambda_i^-=-\lambda_i^+$, and the remaining third corresponds to the unphysical degrees of freedom (e.g. longitudinal polarizations), which have zero Lyapunov exponent, see section \ref{class}. Thus, the classical phase space volume is constant  as long as we don't smear or coarse-grain. As argued in \cite{Kunihiro:2008gv} any measurement provides such a coarse graining due to the quantum mechanical uncertainty relation and thus leads to net entropy growth. For the field theory part of the  $AdS_5/CFT$ dual it thus depends crucially on how precisely the question is asked and how entropy is defined, whether the latter grows or not, i.e. whether information gets lost or not. In this contribution we focus on just one specific detail of this highly complex topic which can be clarified numerically.

\section{Quantum chaos, Lyapunov exponents and the Kolmogorov-Sinai entropy}
Chaos, information scrambling and operator growth in quantum theories can all be studied with the help of  OTOCs of general hermitian operators $V$ and $W$  separated by time $t$:
\begin{equation}
C(t)=-\langle [ W(t), V(0)]^2\rangle_T,
\label{otoc}
\end{equation}
where $\langle \cdot  \rangle_T $ is the thermal expectation value at temperature $T$. By studying the  growth rate of this quantity at times before the Ehrenfest time but well after the dissipation time, we can quantify  "how chaotic" a quantum system is via the exponent $\lambda_{\text{OTOC}}$ in
\begin{equation}
C(t)\sim \hbar^2 e^{2 \lambda_{\text{OTOC}} t}.
\label{def_lambda}
\end{equation}
In \cite{She} it has been famously shown that $ \lambda_{\text{OTOC}}$, given in  natural units, is bounded from above by  $2 \pi$ times the temperature $T$. In addition, large $N$ conformal field theories (CFTs) dual to Einstein gravity are known to saturate this bound
\begin{equation}
 \lambda_{\text{OTOC}}^{CFT,1\ll N}= 2 \pi T.
 \label{lotoc}
\end{equation}
In general, the exponent $ \lambda_{\text{OTOC}}$ as defined in Eq. (\ref{def_lambda}) is going to correspond to the {\it largest} Lyapunov exponent of an entire Lyapunov spectrum (which may be extracted by considering OTCOs of suitable operators) with which the operators $V$ and $W$  overlap. For classical physics Liouville's theorem forces the Lyapunov  spectrum to be symmetric, implying that for every $\lambda_i $ there is a  $\lambda_j$ with $\lambda_j =-\lambda_i$. It is worth noting at this point that, besides the fact that it would appear  natural, there is no proof that this symmetry holds for a quantum version of Lyapunov exponents.  For chaotic quantum systems the authors of \cite{She} determined the exponent $\lambda_{OTOC}$ defined in Eq. (\ref{lotoc})  via an auxiliary function $F(t)$, which decreases with the same exponential rate as $C(t)$ increases. For large $N$ conformal field theories with gravity duals the function $F$ at times $\frac{1}{T}\ll t$ can be written as \cite{She,She2,She3,She4}
\begin{equation}
F(t)= f_0 -\frac{f_1}{N^2}e^{2 \pi T t}+\mathcal{O}(\frac{1}{N^4}),
\label{lmax}
\end{equation}
where $f_0$ and $f_1$ are positive order $\mathcal{O}(1)$ constants depending on the choices for $V$ and $W$. While the operators $V(0)$ and $W(0)$ are  hermitian operators, which can be described as a sum of products, which contain only $\mathcal{O}(1)$ degrees of freedom and the thermal one point functions of $V$ and $W$ should vanish. The  relation (\ref{lmax})  suggests that in this case every positive Lyapunov exponent is maximal $\lambda = 2 \pi T$. Moreover, at first glance it also appears that Eq. (\ref{lmax}) demands that all Lyapunov exponents are positive and that the symmetric structure of the classical Lyapunov spectrum is lost.  However, this point is quite subtle. The question is not whether the commutator of generic operators $W$, $V$ has overlap with an exponentially growing mode, but whether there exist very specific operators, the commutator of which has zero overlap with any such mode. It could have happened that the modest assumptions for $V$ and $W$ that led to Eq. (\ref{lmax}) implied overlap with at least one mode with positive Lyapunov exponent. The idea behind this paper is that, while it might be impossible to extract negative Lyapunov exponents from  Eq. (\ref{lmax}), in practice it is still possible to decide whether they exist by determining the fraction of modes which have positive Lyapunov exponents. In the $N\rightarrow \infty$ and strong coupling limit equation Eq. (\ref{lmax}) suggests that all positive Lyapunov exponents  are expected to be equal  such that, granted Eq. (\ref{entr_gen}) holds,  the ensemble averaged total entropy growth rate is uniquely determined by this fraction. The total entropy growth rate, however, can be calculated as growth rate of the apparent horizon on the gravity side of the duality. Ordering the thoughts above, the following statement can be made: if one is completely agnostic as to what to expect one can imagine three possible scenarios, namely
\begin{itemize}
\item
The Lyapunov spectrum keeps the plus/minus symmetry of classical, chaotic systems, is given by $\pm 2\pi T$, and the relation between Lyapunov exponents and ensemble averaged entropy growth naturally generalizes from statistical mechanics to quantum systems as described by Eq. (\ref{entr_gen}). Then,  the averaged entropy growth rate $\langle d S/ dt \rangle$ is $\pi T$ times  the number of physical degrees of freedom $N_{\text{DOF}}$ (i.e. the total number of Lyapunov exponents).
\item
 Eq. (\ref{entr_gen}) is correct, however there are no modes with negative Lyapunov exponent. Then   $\langle d S/ dt \rangle$ is equal to $ 2 \pi T\cdot N_{\text{DOF}}$.
\item
It is either not true that in this case all Lyapunov exponents  associated with physical degrees of freedom have the maximal absolute value $2\pi T$, or the fraction of negative Lyapunov exponents is some number other than $0$ or $1/2$, or   Eq. (\ref{entr_gen}) is not the correct relation between quantum Lyapunov exponents and and quantum Kolmogorov-Sinai entropy (i.e. state-averaged entropy growth for the quantum system). Then $\langle d S/ dt \rangle/( \pi T N_{\text{DOF}})$ could be any number between $0$ and $2$. 
\end{itemize}
We cannot hope to obtain an indisputable answer, since  $\langle d S/ dt \rangle$ being exactly $ 2 \pi T\cdot N_{\text{DOF}}$ or exactly $  \pi T\cdot N_{\text{DOF}}$ could also be coincidentally a consequence of  the quantum Lyapunov spectrum of large $N$ SYM having some arbitrary shape, that is not $\lambda = \pm 2\pi T$ or $\lambda = 2\pi T$ for all exponents, while e.g. Eq. (\ref{entr_gen}) could be wrong. However, given what we know about the symmetric shape of the classical Lyapunov spectrum (see next section) and its relation to the classical Kolmogorov-Sinai entropy, finding $\langle d S/ dt \rangle \approx \pi T\cdot N_{\text{DOF}}$ would be a strong hint that the quantum Lyapunov exponent is also plus/minus symmetric. 
 \begin{figure}
 \begin{center}
 \includegraphics[scale=0.73]{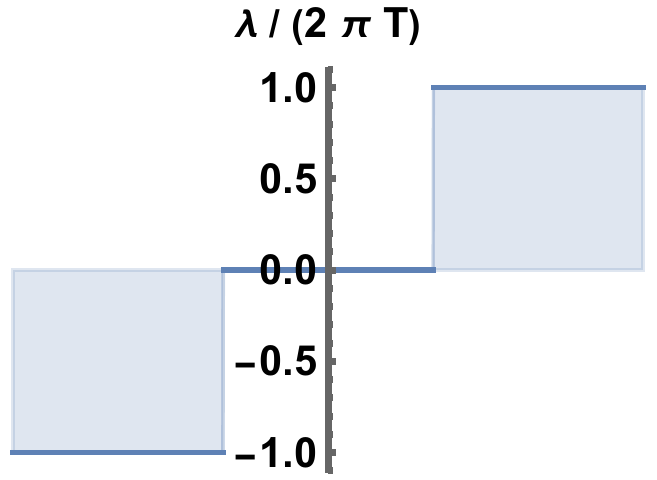}
  \includegraphics[scale=0.73]{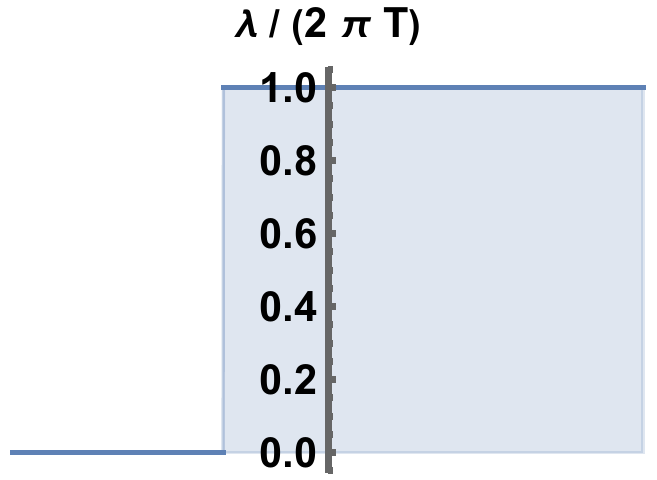}
   \includegraphics[scale=0.73]{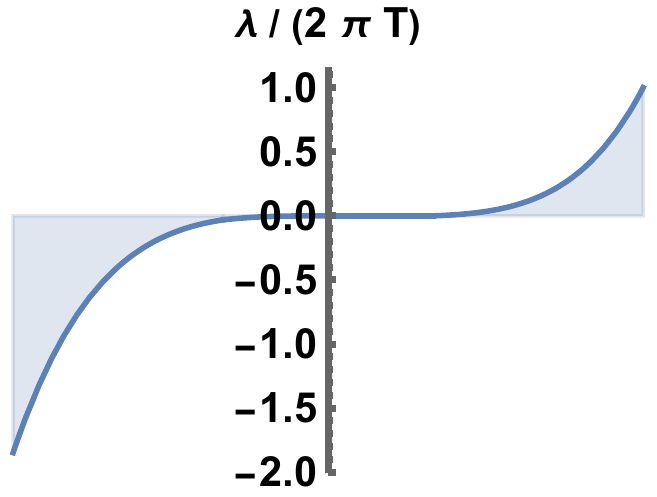} 
 \end{center}
  \caption{Three possible shapes of the Lyapunov spectrum of large $N$, strongly coupled, $\mathcal{N}=4$ SYM in three spatial dimensions, corresponding to the three bullets in the text. In each case the unlabeled x-axis runs over the index $i$ of the Lyapunov exponents $\lambda_i$.  On the left hand side we plot the shape of the spectrum corresponding to the first bullet in the text. All positive Lyapunov exponents are maximal and the Lyapunov spectrum keeps the plus/minus symmetry, such that one half of the Lyapunov exponents corresponding to physical degrees of freedom are $-2 \pi T$. We depict the spectra as potential results that one might obtain from a hypothetical lattice simulation, similar to the one discussed in section \ref{class}, where we  compute  Lyapunov spectra for weakly coupled $SU(2)$ and $SU(4)$ YM theory. Thus, one third of the  exponents is depicted to be vanishing, as they correspond to unphysical degrees of freedom, i.e. longitudinal polarization of gauge fields. For the spectrum on the left the averaged entropy growth rate should fulfill $\langle d S/ dt \rangle/( \pi T N_{\text{DOF}})=1$, where $N_{\text{DOF}}$ denotes the number of physical degrees of freedom. The central plot corresponds to the second bullet, depicting a spectrum where all Lyapunov corresponding to physical degrees of freedom are maximal and thus even the microscopically resolved entropy grows.  Here we have  $\langle d S/ dt \rangle/( \pi T N_{\text{DOF}})=2$. The third plot shows one possible spectrum corresponding to the third bullet. In this case  $\langle d S/ dt \rangle/( \pi T N_{\text{DOF}})=c$ where $c$ can be any number between 0 and 2. Our numerical holographic calculation, discussed in section \ref{results}, strongly suggest that the first figure from left is the most plausible shape.}
 \end{figure}
\section{The Lyapunov spectrum of classical $SU(N)$ Yang Mills Theory}
\label{class}
 As explained in the last section, the central questions to be answered are whether the symmetry between positive and negative Lyapunov exponents persists, i.e. whether Liouville's theorem stays valid, and whether all positive Lypunov exponents approach $\lambda_{\rm max} = 2\pi T$,  assuming that the proposed relation (\ref{entr_gen}) is correct. For these questions some intuition can be gained from studying classical Yang-Mills theories. This is such a natural thing to do and in fact was already done such a long time ago that we do not feel competent to decide who investigated this question first. Instead, we cite the review \cite{Biro:1993qc}. Earlier work can be found there. The rational motivating the study of classical Yang-Mills theory is that  many examples demonstrate that if a classical theory is chaotic the quantized theory is so too, and that many fundamental properties are related (A typical examples are scars in quantum billiards).
  For classical Yang Mills theory it was shown by numerical studies that to high accuracy SU(2) even fullfills the criteria for a globally hyperbolic (Anosov) system \cite{Bolte:1999th}. These criteria concern the dependence of the uncertainty of the numerically obtained Kolmogorov-Sinai entropy on system size and sampling time, a topic we will address below when discussing the precision of our results. These numerical simulations were made by solving the classical Hamilton equations on a finite three dimensional grid, for which one has only a finite number of degrees of freedom, such that it is possible to determine all Lyapunov exponents. Typical results are show in Figs. 9, 10, and 11 of \cite{Biro:1993qc}. One third of the Lyapunov exponents is positive, one third negative, with the same distribution of absolute values, and one third is zero. The latter is due to the fact that a spin 1 field has three degrees of freedom, but for a massless gauge field only two of these are physical while the third is a gauge degree of freedom. These figures show also that even for very small systems, reaching numerically the asymptotic limit, in which all Lyapunov exponents of gauge degrees of freedom are really zero requires very long simulation time. For systems with finite energy density which equilibrate in finite time, such long fitting windows cannot be realized, and the ``intermediate'' Lyapunov exponents of the gauge degrees of freedom are numerically still non-zero, see Figs.~1 and 3 in \cite{Kun}. This shows that the Kolmogorov-Sinai entropy cannot be calculated exactly for finite energy density, which was explored in detail in \cite{Bolte:1999th}. In this contribution we will numerically determine the Kolmogorov-Sinai entropy from the holographic dual, by analyzing the time dependence of the apparent horizon. However, we expect that also in the dual picture the length of time till saturation effects become relevant and the precision with which the slope of the growing apparent horizon area can be determined are related in a similar manner. 

\begin{figure}
\begin{center}
\includegraphics[scale=0.5]{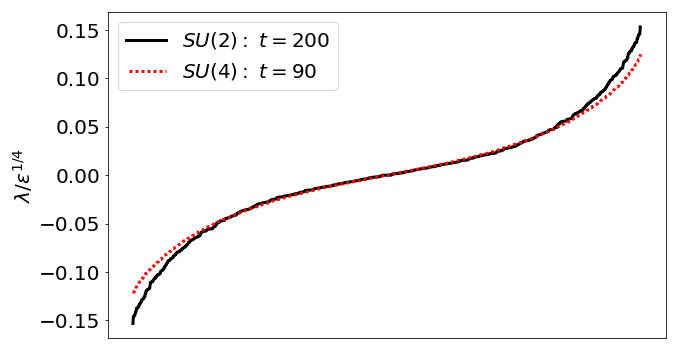}
\end{center}
\caption{ Late time results for the normalized spectrum of Lyapunov exponents for SU(2) and SU(4). The (unlabeled) $x$-axis value parametrizes the index $i$ of the Lyapunov exponent $\lambda_i$.  The feature that one third of all Lyapunov exponents, which correspond to the gauge degrees of freedom, are zero is violated by discretization errors which spoil gauge invariance (see text for more details). The time $t$ is given in units of $\epsilon^{-1/4}$, where $\epsilon$ is the energy density.}
\label{SU2_SU4}
\end{figure}

  Let us comment on a feature of the results obtained in \cite{Kun}, which might otherwise be confusing for a careful reader of that paper: In these numerical calculations space was discretized. The lattice spacing, $\delta$, was tuned to a finite value to obtain the  energy density of the quantum theory also for the classical theory (rather than infinity), i.e. the continuum limit $\delta \rightarrow 0$ was not taken. This finite discretisation implied that spatial derivatives were substituted by quotients of differences, i.e. spatial derivatives became non-local, leading in turn to a violation of local gauge symmetry. This artefacts resulted in the Lyapunov exponents of the gauge degrees of freedom becoming non-zero. This effect is barely visible in Fig.~3 of  \cite{Kun}. To illustrate it more clearly we also performed such an analysis of classical SU(2) and SU(4) theory, see Fig.~\ref{SU2_SU4}. (Time derivatives were not affected which lead in addition to different effects for the $F^{0j}$ (electrical) and $F^{ij}$ (magnetic) components of the field strength tensor.)
  
The symmetry between positive and negative Lyapunov exponents, which is the only point relevant for our discussion, is obviously not affected by this artefact. The bottom line of the discussion in this section is that (semi-) classical\footnote{This calculation is semi-classical in the sense that there is a non-zero $\hslash$ entering the specific choice for the non-vanishing lattice spacing $\delta$.}  YM calculations imply that it would at the very least be an unexpected feature of quantum Lyapunov spectra, if the $\pm$ degeneracy of the spectrum of Lyapunov exponents wouldn't be observed there, too.

\section{Thermalization and holographic isotropization}
\label{isot}
We can simulate an isotropizing, initially far from equilibrium, strongly coupled  SYM plasma, by using the dual gravitational description. Following  the pioneering work of \cite{che1,che2}, there is a large amount of literature on the  topic of studying  out of equilibrium SYM plasmas  via numerical holography (see for example \cite{che3, Fuini:2015hba, wil, therm1, therm2, wilk}), which we cannot do full justice here.  Reference \cite{che3} contains a
 detailed, pedagogical description of the calculation we are going to need in the following. In this section we are going to briefly review  its most important points before continuing to perform ensemble averages over a multitude of isotropization processes. \\ \indent The authors of \cite{che2} studied the numerical evolution of an anisotropic initial state in the CFT, produced by a time dependent shear deformation of the  metric coupling to the stress energy tensor of the CFT. The holographic duality relates states produced by a time dependent, four dimensional metric $h_{\alpha \beta}$ in the large $N$, large 't Hooft coupling CFT to  solutions of five dimensional, classical AdS gravity $g_{\mu \nu}$ in the five dimensional bulk   with the time dependent $4D$ metric as asymptotic boundary. Moreover, the AdS dictionary  allows us to determine the expectation value of the CFT stress energy tensor from the bulk metric
\begin{equation}
\langle T_{\alpha \beta}\rangle = \frac{N^2}{2 \pi^2} \Big(g_{\alpha \beta}^{(2)}-\frac{g^{(2)}_{00} h_{\alpha \beta} }{4} \Big),
\end{equation}
 where $g_{\mu \nu}$ is given in Eddington-Finkelstein coordinates (\ref{inf_coo}) and $g_{\mu \nu}^{(2)}$ represents the second order coefficient of the bulk metric's  expansion around the boundary $\rho=0$.  Assuming spatial homogeneity, this metric ansatz reads
\begin{equation}
g_{\mu \nu}dx^\mu dx^\nu= -\frac{2 dtd\rho}{\rho^2}-2 A(\rho,t)dt^2+\Sigma^2(\rho,t) \hat{g}_{ij}(\rho,t)dx^i dx^j,
\label{inf_coo}
\end{equation}
with $\text{det}(\hat g_{ij})=1$ and near the boundary $\rho \to 0$ one has $A(\rho,t) \sim \frac{1}{2\rho^2}$, $\Sigma(\rho,t) \sim \frac{1}{\rho}$, $\hat g_{ij}(\rho,t) \sim \delta_{ij}$. For simplicity we follow \cite{che3}, where the action on the state by the time dependent, arbitrary shear deformation of the boundary metric is replaced by an arbitrary choice of the anisotropy function $B(\rho,t)$ on the initial Cauchy surface $t=0$, where $B(\rho,t)$ is given via
\begin{equation}
\hat g_{ij}(\rho,t) = \begin{pmatrix}
e^{B(\rho,t)} & 0 & 0\\
0&e^{B(\rho,t)} & 0\\
0&0&e^{-2B(\rho,t)} 
\end{pmatrix}.
\end{equation}
With the ansatz (\ref{inf_coo}) the Einsein equations can be written as a nested system of differential equations on null slices \cite{che3}. In the case of spatial homogeneity, the following data on time slice $t$
\begin{equation}
\{\langle T^{00}(t)\rangle, \,\, \hat g ( \rho,t) \}
\label{data}
\end{equation}
is sufficient to uniquely solve the  system of ordinary differential equations with vanishing spatial gradients, which is most conveniently done using spectral methods \cite{Boyd:Spectral}. The equations of motion of the boundary stress energy tensor 
\begin{equation}
\nabla_\mu \langle T^{\mu \nu}\rangle=0
\end{equation}
together with knowledge of $\hat g (\rho,t)$, $\partial_t \hat g ( \rho,t)-\rho^2 A(\rho,t) \partial_\rho \hat g ( \rho,t)$ and $A(\rho,t)$, the latter two of which are functions we solved the nested system of equations for, allow us to compute the data (\ref{data}) on the next time slice, using fourth order Runge Kutta method, and thus sequentially obtain a solution to the Einstein equations in the bulk. In the setting we consider, one has\footnote{This restricts the  region in Hilbert space from which we draw the  configurations over which we ensemble average. However, this point could also be made, if we performed a similar calculation as in \cite{che3}\iffalse and would work with a time dependent  deformation of the boundary metric and thus a time dependent, but spatially homogeneous energy density\fi. The most general case of an arbitrary time and space dependent boundary metric deformations is numerically not practical, as we are interested in large ensemble averages.  The entropy growth rate for a single configuration at early times can be split into the Kolmogorov-Sinai entropy plus some initial state dependent term $dS/dt = S_{KS}+dS_{i}/dt$, where $S_{KS}$ is constant until thermal equilibrium is approached, while $d S_{i}/dt$ will in general not be constant  and cancels after ensemble averaging. Our ensemble average does yield a constant growth rate for the entropy density $s$, while individual samples to not exhibit this property of the entropy density (see Fig. \ref{bar}). Nonetheless, in future works  it might be interesting to further test our numerical results, by computing (if necessary smaller) ensemble averages of more complicated cases with similar, albeit non-stochastic version of the calculations and numerics described in \cite{am}.  } $\langle T^{00}(t)\rangle=\langle T^{00}(0)\rangle= \frac{3}{8}N^2 \pi^2 T^4$ and a flat boundary metric, while the time dependent pressure components of the stress energy tensor are  anisotropic in the beginning and relax towards the equilibrium value $N^2 \pi^2 T^4/8$
on time scales $t \lesssim 1/T$. A constant energy density  and a non trivial, arbitrary radial dependence of the bulk functions $B(\rho,0)$ (arbitrary up to an appropriate near boundary behavior  $B(\rho,0) \sim \rho^4$ for $\rho \rightarrow 0$) on the initial time slice, can be seen as the result of an appropriate, time dependent, spatially homogeneous deformation of the boundary metric with compact support restricted to $t < 0$,  such that for $0<t$ the asymptotic boundary is  Minkowski space time without deformations and $\langle T^{00}\rangle$ is constant. As in \cite{che3} we use the radial shift invariance of the metric ansatz (\ref{inf_coo}) to keep the apparent horizon at a constant $ \rho_h \pi T =1$, where $T$ refers to the equilibrium temperature. We use the scaling symmetry $\rho \to \alpha \rho$ and $x_i \to\alpha x_i$ to set $\rho_h=1$ throughout our numerical simulations. Physical quantities are given in units of \begin{equation}
\mu = \pi T .
\end{equation}
 \section{Results}
 \label{results}
Our aim is to compute the   Komogorov-Sinai entropy for $\mathcal{N}=4$ SYM in the large $N$ limit using holography. We consider an ensemble of out of equilibrium, spatially homogeneous but anisotropic states, that evolve towards the same thermal state, in strongly coupled $\mathcal{N}=4$ SYM. Different initial anisotropic states correspond to different choices for the anisotropy function $B$ introduced in section \ref{isot}. Using the holographic principle, we can determine the number of physical degrees of freedom (DOF) per unit volume of the SYM plasma\footnote{When we henceforth refer to 'Lyapunov exponents' we explicitly exclude those, that correspond to unphysical degrees of  freedom, which, e.g., one has to deal with in lattice simulations (see section (\ref{class})). }. Once the plasma has reached thermal equilibrium the dual description is a Schwarzschild black  hole geometry and its number of DOF $N_\text{DOF}$ is usually taken to be its area measured in Planck length squared\footnote{To the best of our knowledge this relation is still unproven for Schwarzschild black holes.}. Thus $N_\text{DOF}$ can be computed from the  Bekenstein-Hawking entropy of the black  hole, such that the number of DOF per unit volume is
\begin{equation}
 \frac{N_\text{DOF}}{V}= \frac{N^2 \mu^3}{2 \pi}.
\end{equation}
This implies that the   Kolmogorov-Sinai entropy density $s_{KS} = S_{KS}/V$ should be given by
\begin{equation}
s_{KS} = \bigg\langle\frac{d s}{dt} \bigg\rangle = \frac{1}{V} \sum_{\lambda >0 }\lambda = c\, \frac{N^2}{2} \pi^3 T^4.
\label{sks}
\end{equation}
Where $c$ in (\ref{sks}) is equal to 2, if every Lyapunov exponent is positive and maximal, and equal to 1, if every positive Lyapunov exponent is maximal and the Lyapunov spectrum stays plus/minus symmetric. This is assuming that the averaged entropy growth rate of black holes actually saturates the upper bound, derived from the upper bound on quantum Lyapunov exponents. We can compute $s_{KS}$ via holography by ensemble averaging over the growth rates of the apparent horizons' volume element, were the ensemble consists of numerical simulations of the isotropizing SYM plasma, described in section \ref{isot}, with different initial anisotropy functions. Let $g(\rho_h,t)$ be the determinant of the  metric induced on the apparent horizon on timeslice $t$, then the CFT entropy density grows as
\begin{equation}
\frac{d s}{dt} =\frac{N^2}{2} \pi^3 T^4  \frac{d}{d\,\mu t} \sqrt{g(\rho_h,t)}=\frac{N^2}{2} \pi^3 T^4   \frac{d}{d\,\mu t} \Sigma(\rho_h,t)^3,
\label{sks_hol}
\end{equation}
with $t$ given in units of $\mu^{-1}$. From equating  (\ref{sks}) and the ensemble averaged version of (\ref{sks_hol}) we can extract the Kolmogorov-Sinai entropy and the coefficient $c$. The main challenge is to find 'good' ensembles of states $\{\phi\}$, which sample sufficiently large parts of the Hilbert space\footnote{With the constraint that we are only interested in states that evolve to the same thermal state.} without any bias, such that
\begin{equation}
\Big \langle \frac{ds}{dt} \Big \rangle_{\{\phi\}} \approx  \Big \langle \frac{ds}{dt} \Big \rangle,
\end{equation}
where $\langle \cdot \rangle$ is the state average described in the introduction, section \ref{intro}, and $\langle \cdot \rangle_{\{\phi\}}$ is the average over the ensemble ${\{\phi\}}$.\footnote{Rigorously showing that an ensemble is 'good' in the sense described above appears close to impossible. Necessary requirements on ensembles include that known Hilbert space averages should be matched by the approximation  $\langle \cdot \rangle_{\{\phi\}}$ to good accuracy. In this light we checked that for the ensembles we studied one has that $\langle T_{ij} \rangle_{\{\phi\}}$ matches the equilibrium value $N^2 \pi^2 T^4/8$ up to a small error. For instance in the case of ensemble (IV) we find a maximal deviation from the equilibrium value of $1.6\%$.}
  We generate multiple ensembles by choosing $\phi \in  C^{\infty}([0, 1])$   randomly with \begin{equation}
  B(\rho,0)=\rho^4 \phi(\rho),
  \end{equation}
 while the set of different ensembles we consider can be classified into two main categories: On the one hand we choose $\phi(\rho)$ from finite dimensional subspaces of the function space $  C^{\infty}([0, 1])$ generated by polynomials and Gaussians with random coefficients. On the other hand we generate random points for each element of an equidistant grid on $ [0, 1]$, interpolate and smooth the resulting function  (we smooth via filtering out large radial derivatives in order to improve numerical stability). In total we collected data from $5$ different ensembles ranging in size from  several hundred thousand to 3 million simulations. Ensemble (I) was generated by choosing $N_p$ random real numbers $r_i$ between $-15$ and $15$, while $N_p$ itself is a random integer between $5$ and $199$. The resulting list of points $\{i/N_p,r_i\}$ is filtered with a Gauss-filter of width $3/N_p$. The random function $\phi$ is then found by fitting the filtered list with a polynomial of order 15. Ensemble (II) was generated from  initial anisotropy functions of the form
\begin{equation}
B(\rho,0)= \rho^4 \bigg(\beta_1 \exp{\Big(-\frac{(\rho-\rho_1)^2}{w_1^2}\Big)} +\beta_2 \exp{\Big(-\frac{(\rho-\rho_2)^2}{w_2^2}\Big)}+a_1(a_2-\rho)^5+a_0 \bigg),
\end{equation}
with $\beta_{1,2} \in [-10,10]$, $w_{1,2} \in [-5,5]$, $\rho_{1,2} \in [-0.5,0.5]$, $a_{0,1} \in [-4,4]$, $a_2 \in [-1,1]$ drawn from uniform distributions.  Ensemble (III) is generated analogously to ensemble (II), but with 3 instead of 2 Gaussians, where the corresponding parameters $\rho_3$, $\beta_3$ and $w_3$ have the same range as $\rho_{1,2}$, $\beta_{1,2}$ and $w_{1,2}$. For ensemble (IV) we chose $N_p$ random points between $-5$ and $5$, where $N_p$ is again a random integer between $5$ and $99$. We then apply a low pass filter onto the list of random points and transform directly from the equidistant grid to a Chebyshev grid via spectral methods. Finally ensemble (V) is  generated analogously to (I), but  now also the order of the interpolating polynomial is randomly chosen between 5 and 25 and the Gauss-filter has width $4/N_p$. In each case we represent $\phi(\rho)$ as a vector of $26$ values on a Chebyshev grid and solve the system of differential equations using spectral methods. 
\begin{figure}
\begin{center}
\includegraphics[scale=0.3]{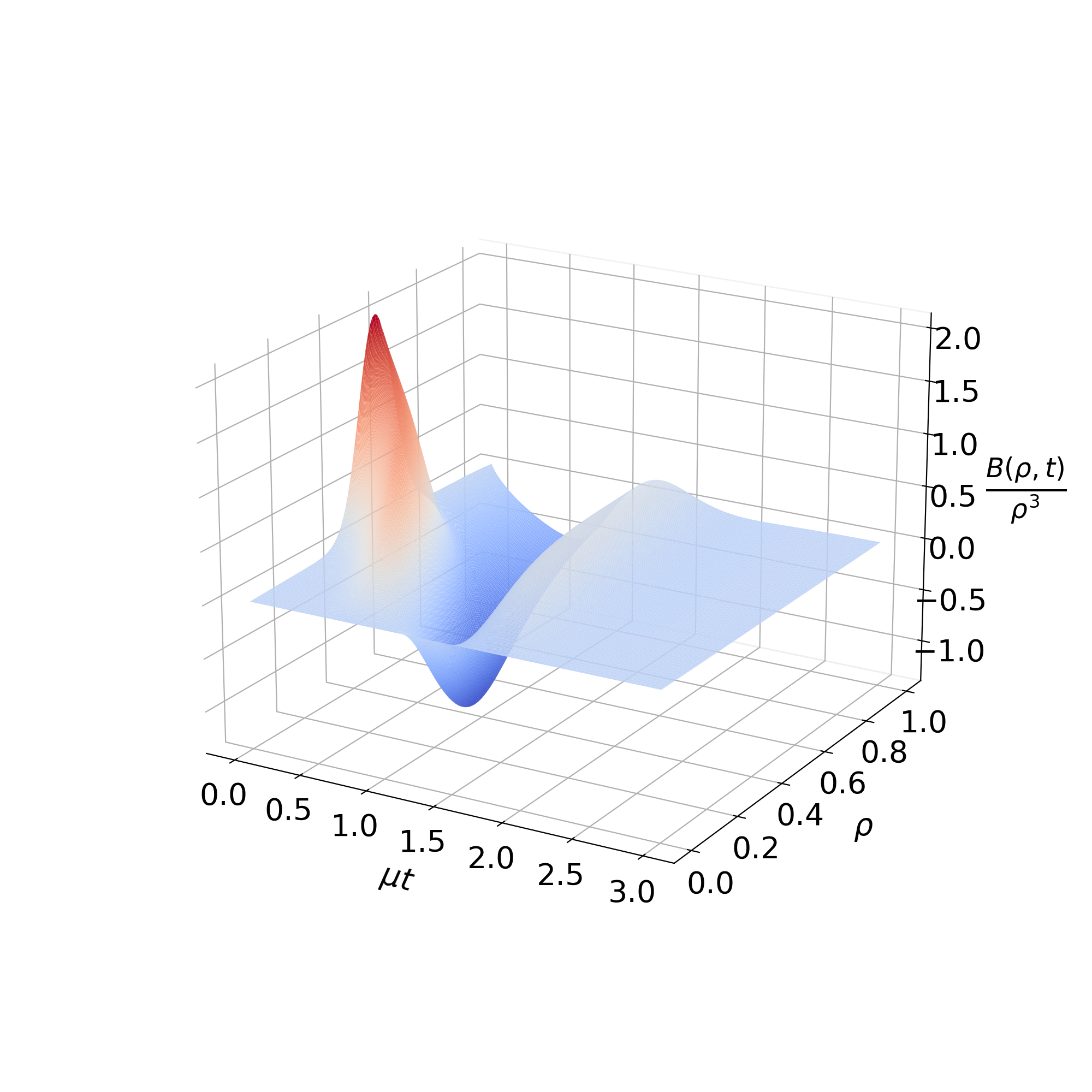}
\includegraphics[scale=0.3]{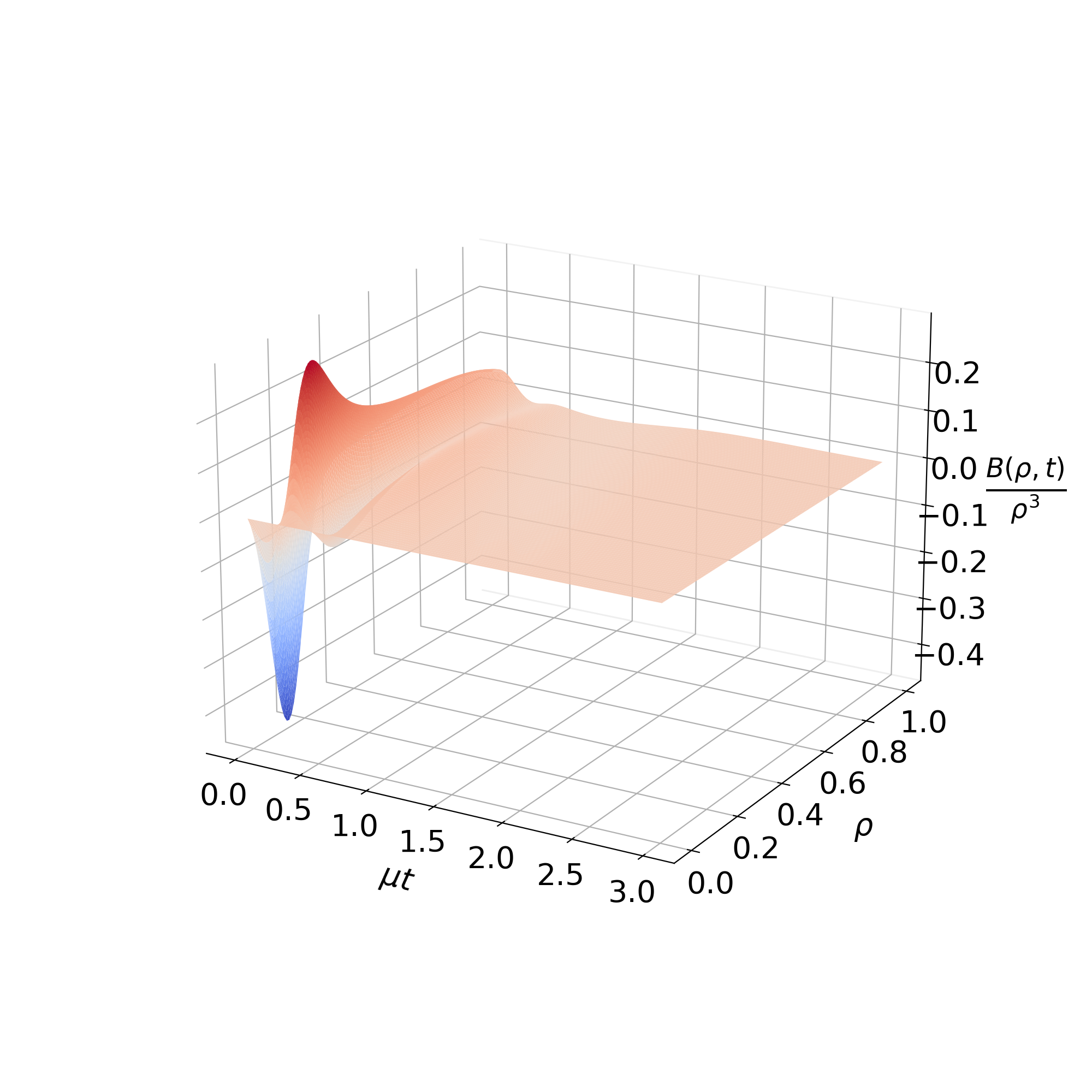}
\end{center}
\caption{The rescaled anisotropy function $B(\rho,t)/\rho^3$ for two random
 samples from our ensemble (II). The radial coordinate $\rho$ is shifted, such that the apparent horizon is positioned at $\rho=1$.}
\end{figure}
\\ \indent For  classical, chaotic systems   we know what one might expect regarding the behavior of the ensemble averaged entropy. There the entropy $S(t)$ follows a general pattern (see \cite{bar}). In the first, short stage (1) the behavior of $S(t)$  is dominated by
the initial distributions and no  general statement can be made\footnote{The larger the ensemble average, the less $S(t)$ depends on random initial configurations and the shorter this first period should be.}. In the second stage  (2) $S(t)$ grows approximately linearly and the growth rate corresponds to the Kolmogorov-Sinai entropy, the sum over all positive Lyapunov exponents. Then in the third stage (3) the entropy tends asymptotically towards its equilibrium value\footnote{In our case thermal equilibrium is synonymous with $ \Sigma(\rho_h,t)=1$.}. However, for individual runs with inconvenient choices of initial conditions (i.e. initial configurations that already start close to thermal equilibrium, or for which  the entropy grows so fast in the first stage that the system is brought close to equilibrium already there)  stage (1) and stage (3) might merge,
skipping stage (2) in which we are interested. To avoid these pathological contributions to our ensemble average, we consider initial conditions which are far from equilibrium (we both consider averages of runs for which $\Sigma^3(\rho_h,t=0) = 0.1\pm 0.01$ and $\Sigma^3(\rho_h,t=0) = 0.01\pm 0.01$)  and focus on initial configurations for which the entropy density at the starting time $t=0$ doesn't grow faster than the threshold $N^2 \pi^3 T^4 a/2$, while we display our results as a function of the threshold or cut-off parameter $a$ in Fig. \ref{fa}. We indicate averages over runs for which the slope of the entropy density $\partial_t s|_{t=0}$ at initial time $t=0$   does not exceed\footnote{Or, put differently, for which $\frac{d}{d\mu t} \Sigma^3(\rho_h,t)\Big|_{t=0} < a$.}  $N^2 \pi^3 T^4 a/2$ with a subscript $a$ by $\langle \cdot \rangle_a$. We find that the growth rate of the ensemble averaged entropy density  during the time period in which it grows  linearly does not depend on the cut-off parameter $a$ for   a wide variety of different cut-off choices $a \in [0,5]$.  For  large cut-off values $a > 5$  pathological contributions of the form
as
 shown in Fig. \ref{mt10}, which skip stage (2), have non-negligible influence on some ensemble averages. See caption of Fig. \ref{fa} for more details.
\begin{figure}
\begin{center}
\includegraphics[scale=0.6]{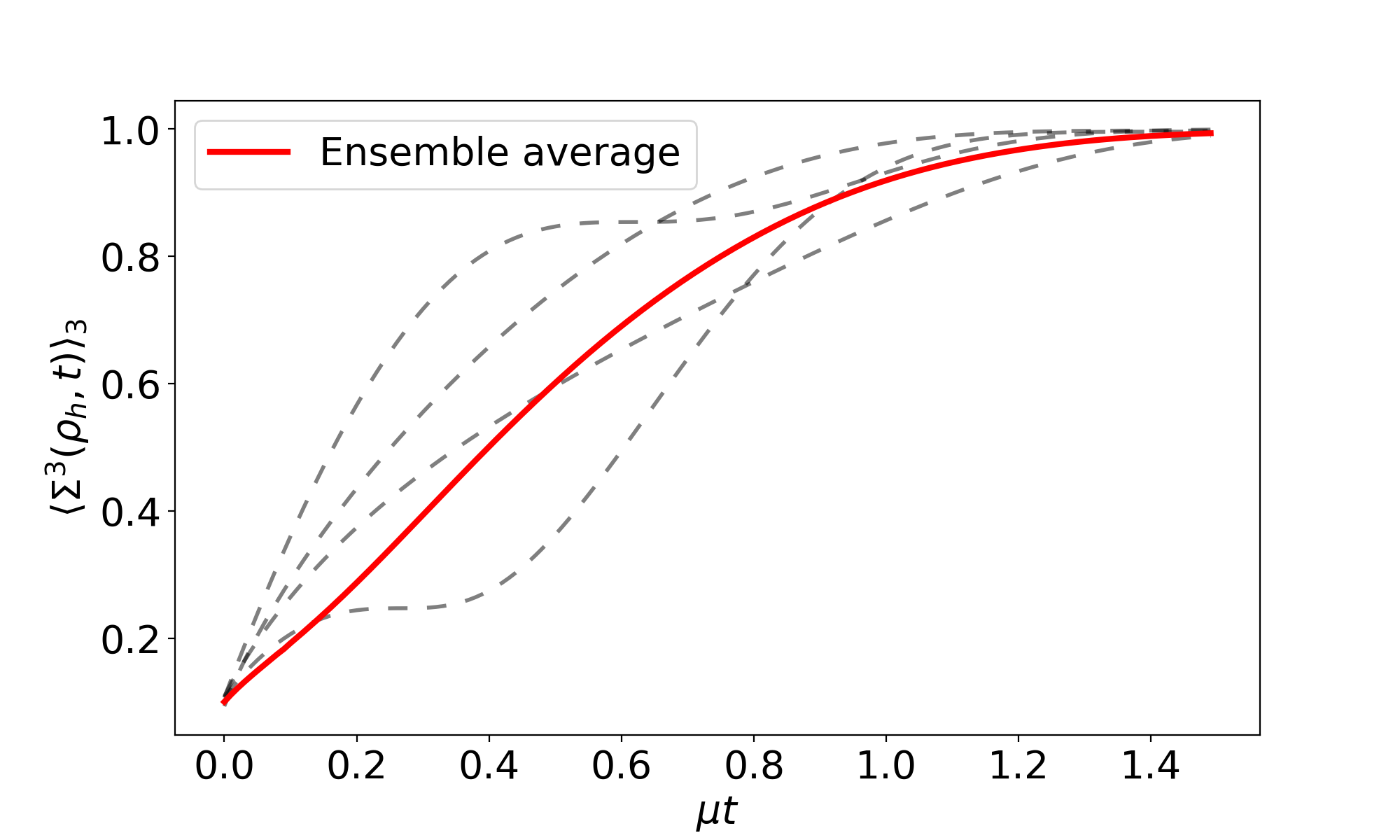}
\end{center}
\caption{We show $\langle \sqrt{g(\rho_h,t)}\rangle_3=\langle \Sigma(\rho_h,t)^3\rangle_3$ as a function of time $\mu t$, which corresponds to the ensemble (II) averaged CFT entropy density  in units of $\frac{N^2}{2 \pi} \mu^3$ with cut-off parameter $a=3$. For the ensemble average  (red curve) we obtain a constant slope in the interval $\mu t \in [0.05,0.5]$. The grey dashed curves show the corresponding plots for a selection of single simulations in our ensemble. Our results only depend negligibly on the cut-off $a$  (see Fig. \ref{fa}) as long as $a$ is chosen within $a \in [0,5]$, such that contributions with very large initial slopes, which skip stage (2), are suppressed.}
\label{bar}
\end{figure}
\begin{figure}
\begin{center}
\includegraphics[scale=0.6]{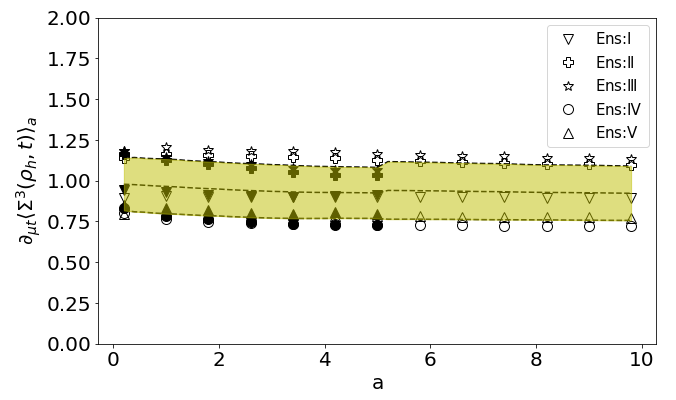}
\end{center}
\caption{Here we show the constant slope during the linear growth phase of the ensemble averaged area element of the apparent horizon $\langle \Sigma(\rho_h,t)^3\rangle_a$, the dual of the CFT entropy, in units of $\frac{N^2}{2 \pi} \mu^4$ as a function of the cutting threshold $a$. The subscript $a$ in $\langle \cdot \rangle_a$ indicates that the ensemble average is taken over all histories, for which $\frac{d}{d\mu t} \Sigma(\rho_h,t)^3\big |_{t=0}$ does not exceed $a$. The results displayed above are computed for  averages over simulations from our 5 different ensembles described in the text, which all start far from equilibrium, i.e. $\Sigma(r_h,t=0)^3 \approx 0.1$ (filled symbols) or $\Sigma(r_h,t=0)^3 \approx 0.01$ (empty symbols). For each ensemble and each value of $a$ we determine a linear fit for the averaged entropy growth. The results of the slopes of those fits are displayed above. The average of our results is very close to $1$ (center dashed line) with variance $\approx 0.16$.  For  very large initial growth rates $a$ the entropy jumps close to equilibrium within a very short initial time span $t < 1/(a\mu)$\iffalse, even for $\Sigma(r_h,t=0)^3 \approx 0.1$\fi. Stage (2) is skipped in this case and a linear growth cannot be observed.  Thus, for  $a>5$  some ensembles averages (mainly those starting at $\Sigma(r_h,t=0)^3 \approx 0.1$) get 'spoiled' by contributions to the average similar to those depicted in Fig. \ref{mt10}.  \iffalse developed code gives always a result, even if a curve is obviously dominated by artefacts, as e.g. shown in Fig \ref{mt10}. This explains  The vertical bars represent the errors of the linear fit, the length of the fit-range was chosen to be $0.6/\mu$. For larger cut-off values $a>4$ the errors increase while the estimated growth rate of  $\langle \Sigma(\rho_h,t)^3\rangle_a$ decreases. This behavior is explained by  the ensemble average being increasingly influenced by runs with large initial slopes, such that they skip stage (2) and are nowhere linear (see e.g. Fig. \ref{mt10}). Thus the growth rate computed from linear fits to ensemble averages with large cut-off parameters are not reliable. Much like the decreasing slope at large $a$, the jump at about $a \approx 7$ in the plot above is unphysical and explained by the fitting algorithm selecting a new $t$-interval for which the error to the linear fit is smaller with the selected fit-range. \iffalse For processes with very large initial slopes (e.g. $\partial_t\langle \sqrt{|g(r_h,t=0)|}\rangle>6$), or processes which start already close to thermal equilibrium stage (2) is skipped and the Kolmogorov-Sinai entropy cannot be read off from the  entropy growth rate. \fi \fi}
\label{fa}
\end{figure}
\begin{figure}
\begin{center}
\includegraphics[scale=0.6]{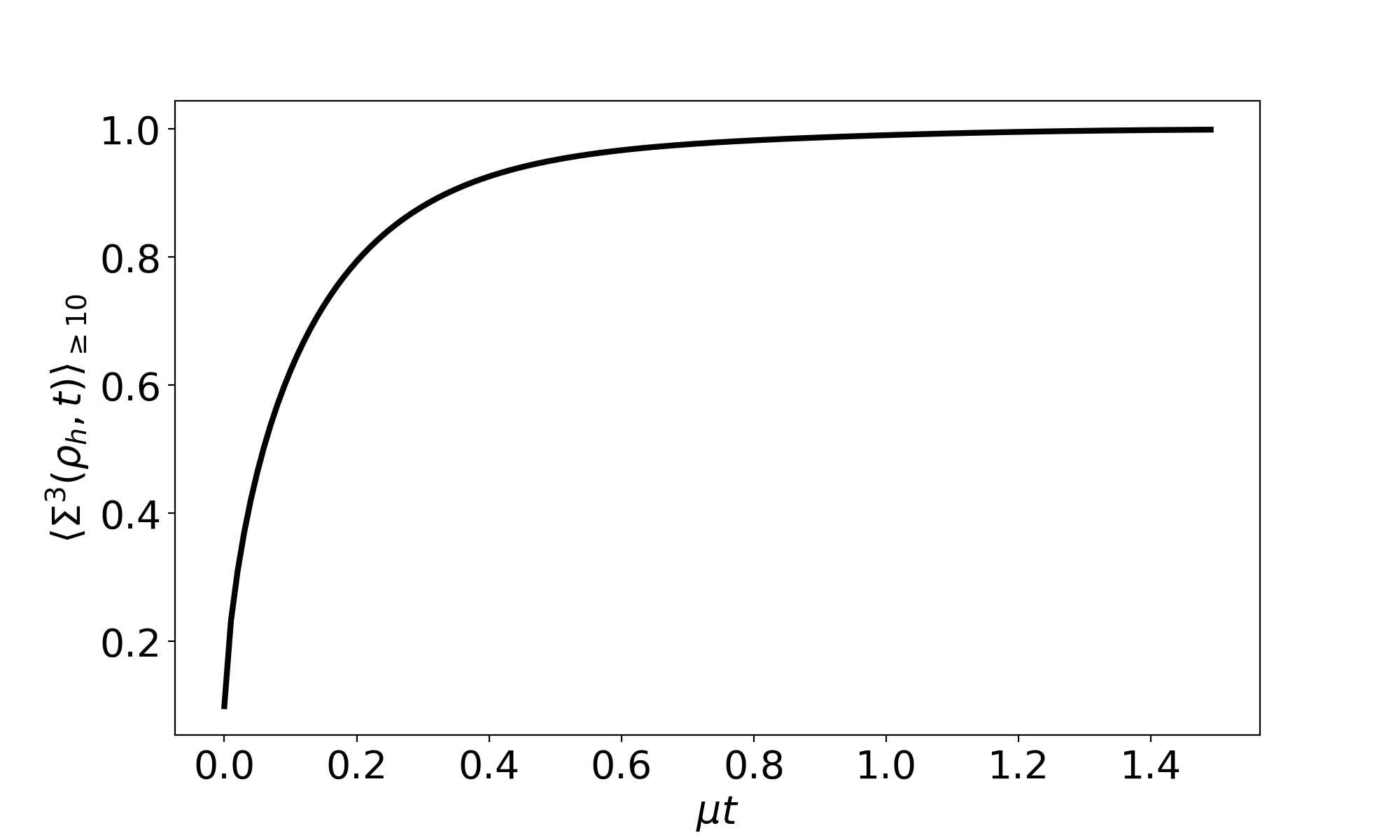}
\end{center}
\caption{The averaged horizon area per boundary volume in units of $\frac{N^2}{2 \pi} \mu^3$. Here we exclusively average over those simulations with very large initial growth rates of the the entropy density, specifically for which $10\leq \frac{d}{d\mu t} \Sigma(\rho_h,t)|_{t=0}$. The initial slope in this example is so large, that the system reaches near-equilibrium before the linear growth phase can start. The results shown above correspond to ensemble (II).}
\label{mt10}
\end{figure}
We find 
\begin{equation}
  \frac{d}{d\mu t} \big \langle \Sigma(\rho_h,t)^3 \big \rangle_{a =0} = 0.98 \pm 0.16.
  \label{res}
\end{equation}
The large error is owed to our ignorance of which type of ensembles gives the best approximation to the average over all states in Hilbert space that evolve towards the same thermal state with temperature $T$. Nonetheless it is 
 intriguing that we find a result of approximately 1 in (\ref{res}), which neatly fits to the physical intuition, that the ensemble averaged entropy growth rate for AdS/CFT saturates the theoretical maximal value (\ref{entr_gen}), the Lyapunov spectrum of $\mathcal{N}=4$ SYM inherits the $\pm$-symmetric structure of the classical YM theory (see section \ref{class}) and that in the holographic limit all positive Lyapunov exponents are maximal.
 The heuristic explanation  for the symmetry of the Lyapunov spectrum of classical YM theory is its time reversal invariance. This is equivalent to a constant microscopically resolved, fine grained entropy, i.e. a constant phase space volume, which implies that for every direction in phase space, in which the phase space volume grows with rate $e^{\lambda t}$ there has to  be another direction in which it contracts with rate $e^{-\lambda t}$. Thus, one could make a point that any reasonable\footnote{Reasonable in the sense that the classical Lyapunov spectrum is obtained in the classical limit.} generalization of the classical Lyapunov spectrum to a quantum Lyapunov spectrum should inherit this symmetry as long as the quantum theory is unitary. 
\FloatBarrier
 \section{Conclusion}
 By ensemble averaging over a multitude of isotropization processes we found that the Kolmogorov-Sinai entropy density of large $N$, $\mathcal{N}=4$ SYM at strong coupling is given by
 \begin{equation}
 s_{KS} \approx \frac{N^2 \pi^3 T^4}{2}.
 \label{sk_res}
 \end{equation}
 We argued  that the two most plausible shapes of the Lyapunov spectrum of strongly coupled, large $N$ CFTs  with Einstein gravity duals (such that the MSS-bound is saturated) is either all Lyapunov exponents being positive and maximal $\lambda = 2\pi T$, or a degenerate spectrum $\lambda = \pm 2\pi T$, such that $\sum_\lambda =0$, with the Lyapunov spectrum keeping the $\pm$ symmetry that we are used to in the case of classical YM theory, or classical physics in general. In the case of a degenerate spectrum, the result (\ref{sk_res}) implies that the intuition, that  large $N$, $\mathcal{N}=4$ SYM at strong coupling has the largest possible Kolmogorov-Sinai entropy fulfilling (during the linear growth phase of the ensemble averaged entropy)
\begin{equation}
 S_{KS}= \Big \langle \frac{d S}{dt}  \Big \rangle = \sum_{\lambda > 0} 2 \pi T,
\end{equation}
appears to be correct. \\ \indent  One interesting statement derived from AdS/CFT is that the quark gluon plasma, produced during heavy ion collisions, thermalizes very quickly \cite{che3}  on time scales that are just a fraction of $1 fm/c$. Even when finite 't Hooft coupling corrections are taken into  account \cite{wae1, wae2} or non-trivial transverse fluctuations of the energy density are considered \cite{ wae3, mue}, both of which roughly doubling the thermalization time, one still ends up with a result that is  below $1 fm/c$, which does not contradict experimental observations, but rather estimates from weakly coupled, $N=3$ YM-calculations \cite{Kun}. Granted that QCD at high temperatures strongly  resembles large $N$, $\mathcal{N}=4$ SYM, which  according to (\ref{sk_res}) is likely to actually saturate the possible upper bound on  the (ensemble averaged) entropy production rate, this mismatch between weak coupling results on the one side and  phenomenology and holography on the other side
is not surprising.  \\
In future works we will further test the results obtained in this paper by considering simulations with non-homogeneous initial conditions and arbitrary boundary metrics. Other interesting questions to explore in  this context are how the Kolmogorov-Sinai entropy behaves at finite coupling, or how/whether the situation changes, when we replace the entropy computed via the apparent horizon area by the entanglement entropy of some boundary region, weighted by the measure of this boundary area.
 \section{Acknowledgements}
 We thank Larry Yaffe, Amos Yarom, Berndt M\"uller, Ben Meiring and Masanori Hanada for helpful comments and discussions. 
 SW acknowledges support in part 
by an Israeli Science Foundation excellence center grant 2289/18, a Binational Science
Foundation grant 2016324, the U.S. Department of Energy grant  DE-SC-0011637 and the Feodor-Lynen fellowship program of the Alexander von Humboldt foundation. AS was supported by BMBF (Project 05P2018, grant number 05P18WRCA1).

\FloatBarrier

\end{document}